\documentclass[keeplastbox]{article}
\usepackage{spconf,amsmath,graphicx}
\usepackage{amsfonts}
\usepackage{lipsum}
\usepackage{mathtools}
\usepackage{cuted}
\usepackage{flushend}
\usepackage{tikz}
\usepackage{adjustbox}
\usepackage{setspace}
\usetikzlibrary{arrows.meta,spy}
\usepackage{caption}
\usepackage{etoolbox}

\patchcmd{\thebibliography}
  {\settowidth}
  {\setlength{\itemsep}{0pt plus 0.1pt}\settowidth}
  {}{}

\title{Combining Shape Priors with Conditional Adversarial Networks for Improved Scapula Segmentation in MR images}
 \name{A. Boutillon$^{1,2}$ \qquad B. Borotikar$^{2,3,4}$ \qquad V. Burdin$^{1,2}$ \qquad P.-H. Conze$^{1,2}$ }

\address{$^{1}$ IMT Atlantique, Brest, France \\
         $^{2}$ LaTIM UMR 1101, Inserm, Brest, France \\
         $^{3}$ Universit\'e de Bretagne Occidentale (UBO), Brest, France\\
         $^{4}$ Centre Hospitalier R\'egional et Universitaire (CHRU) de Brest, Brest, France}
\begin{document}
%
\maketitle
\begin{abstract}
This paper proposes an automatic method for scapula bone segmentation from Magnetic Resonance (MR) images using deep learning. The purpose of this work is to incorporate anatomical priors into a conditional adversarial framework, given a limited amount of heterogeneous annotated images. Our approach encourages the segmentation model to follow the global anatomical properties of the underlying anatomy through a learnt non-linear shape representation while the adversarial contribution refines the model by promoting realistic delineations. These contributions are evaluated on a dataset of 15 pediatric shoulder examinations, and compared to state-of-the-art architectures including UNet and recent derivatives. The significant improvements achieved bring new perspectives for the pre-operative management of musculo-skeletal diseases.

\end{abstract}
\vspace{-.5em}
\section{Introduction}
\label{sec:intro}

Semantic segmentation is a crucial supportive technology for the entire clinical imaging workflow, which helps diagnose patient-specific treatement strategies through 3D models and simulations. Manual segmentation is typically a tedious, time-consuming process which is often prone to intra and inter-expert variability. Interest in developing fast and automatic approaches recently arose to help clinicians for diagnosing pathologies, planning therapeutic interventions and predicting interventional outcomes. Given the success of deep learning for natural image processing, the scientific community is keen on adopting such a methodology for medical imaging with promising performance, especially for segmentation issues. In the medical community, the most well-known architectures for segmentation are UNet \cite{ronneberger_u-net:_2015} and its 3D counterpart VNet \cite{milletari_v-net:_2016}. Without any hand-crafted features, these methods yield impressive results compared to traditional approaches. Consequently, extensions have been proposed to further improve the quality of delineations.


\begin {figure}
\centering
\begin{adjustbox}{width=.48\textwidth}
\tikzstyle{dashed}=[dash pattern=on .9pt off .9pt]

\begin{tikzpicture}

\node[inner sep=0pt] (mri) at (0,0)
    {\includegraphics[width=.02\textwidth]{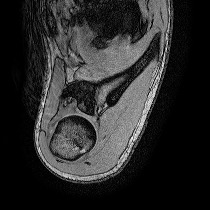}};
\node[inner sep=0pt] (pred) at (1.3,0)
    {\includegraphics[width=.02\textwidth]{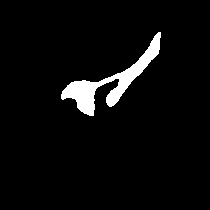}};
\node[inner sep=0pt] (gt) at (1.3,-.8)
    {\includegraphics[width=.02\textwidth]{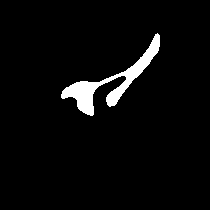}};

\draw[line width=0.01mm, fill=cyan!30] (.35,.25) -- (.65,.125) -- (.95,.25) -- (.95,-.25) -- (.65,-.125) -- (.35,-.25) -- cycle;
\node at (.65,0) {\fontsize{2.3}{2.3}\selectfont UNet};

\draw[line width=0.01mm, fill=orange!40] (1.65,.25) -- (1.95,.125) -- (1.95,-.125) -- (1.65,-.25) -- cycle;
\draw[line width=0.01mm, fill=orange!40] (1.65,-.55) -- (1.95,-.675) -- (1.95,-.925) -- (1.65,-1.05) -- cycle;
\node at (1.8,0) {\fontsize{2.3}{2.3}\selectfont Encoder};
\node at (1.8, -.8) {\fontsize{2.3}{2.3}\selectfont Encoder};

\draw[line width=0.01mm, fill=red!40] (1.05,.38) -- (1.55,.38) -- (1.425,.68) -- (1.175,.68) -- cycle;
\node at (1.3, .53) {\fontsize{1.8}{1.8}\selectfont Discriminator};

\draw[line width=0.01mm] (1.975,-.35) rectangle (2.225,-.45) node[pos=.5] {\fontsize{2.3}{2.3}\selectfont L2};
\draw[line width=0.01mm] (1.425,-.35) rectangle (1.175,-.45)  node[pos=.5] {\fontsize{2.3}{2.3}\selectfont Dice};
\draw[line width=0.01mm] (1.425,.68) rectangle (1.175,.78) node[pos=.5] {\fontsize{2.3}{2.3}\selectfont BCE};

\draw[line width=0.01mm, -{Latex[length=1.5pt, width=1.5pt]}] (mri.east) -- (.35,0);
\draw[line width=0.01mm, -{Latex[length=1.5pt, width=1.5pt]}] (.95,0) -- (pred.west);
\draw[line width=0.01mm, -{Latex[length=1.5pt, width=1.5pt]}] (pred.east) -- (1.65,0);
\draw[line width=0.01mm] (pred.south) -- (1.3,-.2);
\draw[line width=0.01mm, -{Latex[length=1.5pt, width=1.5pt]}] (1.3,-.27) -- (1.3,-.35);
\draw[line width=0.01mm, -{Latex[length=1.5pt, width=1.5pt]}] (gt.north) -- (1.3,-.45);
\draw[line width=0.01mm, -{Latex[length=1.5pt, width=1.5pt]}] (gt.east) -- (1.65,-.8);

\draw[line width=0.01mm] (1.95,0) -- (2.1,0);
\draw[line width=0.01mm, -{Latex[length=1.5pt, width=1.5pt]}] (2.1,0) -- (2.1, -.35);

\draw[line width=0.01mm] (1.95,-.8) -- (2.1, -.8);
\draw[line width=0.01mm, -{Latex[length=1.5pt, width=1.5pt]}] (2.1, -.8) -- (2.1,-.45);

\draw[line width=0.01mm, dashed, -{Latex[length=1.5pt, width=1.5pt]}] (2.225, -0.4) -- (2.275, -0.4) -- (2.275,-1.1) -- (.65,-1.1) -- (.65,-.475);

\draw[line width=0.01mm, dashed, -{Latex[length=1.5pt, width=1.5pt]}] (1.175, -0.4) -- (.725, -0.4);

\draw[line width=0.01mm, dashed, -{Latex[length=1.5pt, width=1.5pt]}] (1.3,.78) -- (1.3,.83) -- (-.25,.83) -- (-.25,-.4) -- (.575,-.4);

\draw[line width=0.01mm,dashed, -{Latex[length=1.5pt, width=1.5pt]}] (.65, -.325) -- (.65, -.125);

\draw[line width=0.01mm] (.65,-.4) circle (.075);
\draw[line width=0.01mm] (.65,-.35) -- (.65,-.45);
\draw[line width=0.01mm] (.6,-.4) -- (.7,-.4);

\draw[line width=0.01mm, -{Latex[length=1.5pt, width=1.5pt]}] (1.4,.18) -- (1.4,.38);
\draw[line width=0.01mm] (0,.18) -- (0,.28) -- (1.2,.28);
\draw[line width=0.01mm, -{Latex[length=1.5pt, width=1.5pt]}] (1.2,.28) -- (1.2,.38) ;

\node at (0,-.23) {\fontsize{2.5}{2.5}\selectfont MRI};
\node at (1.3,-.23) {\fontsize{2.5}{2.5}\selectfont Prediction};
\node at (1.3,-1.03) {\fontsize{2.5}{2.5}\selectfont Groundtruth};

\draw[line width=0.01mm, -{Latex[length=1.5pt, width=1.5pt]}] (-.2,-.75) -- (0, -.75);
\draw[line width=0.01mm, dashed, -{Latex[length=1.5pt, width=1.5pt]}] (-.2,-1) -- (0, -1);

\node[anchor=west] at (-.1, -.7) {\fontsize{2.5}{2.5}\selectfont Forward };
\node[anchor=west] at (-.1, -.8) {\fontsize{2.5}{2.5}\selectfont propagation};
\node[anchor=west] at (-.1, -.95) {\fontsize{2.5}{2.5}\selectfont Backward };
\node[anchor=west] at (-.1, -1.05) {\fontsize{2.5}{2.5}\selectfont propagation};

\end{tikzpicture}
\end{adjustbox}
\caption{Proposed training framework based on UNet \cite{ronneberger_u-net:_2015} exploiting Dice, Binary Cross Entropy (BCE) and latent euclidean (L2) losses.}
\label{fig:fig_framework}
\end{figure}
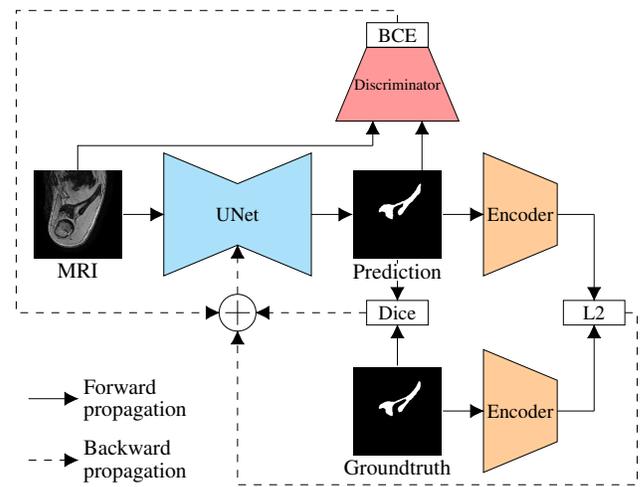



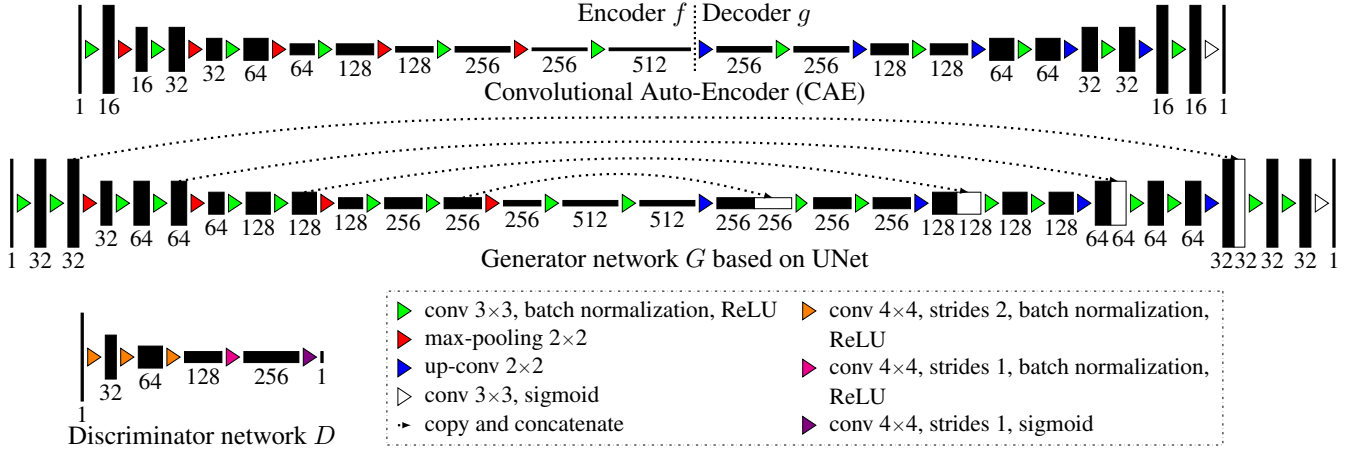
\begin {figure*}
\centering
\begin{adjustbox}{width=\textwidth}

\begin{tikzpicture}

\draw[line width=0.01mm, fill=black] (-.1,-2) rectangle (0,2);
\draw[line width=0.01mm, fill=black] (1,-2) rectangle (1.5,2);
\draw[line width=0.01mm, fill=black] (2.5,-2) rectangle (3,2);

\draw[line width=0.01mm, fill=black] (4,-1) rectangle (4.5,1);
\draw[line width=0.01mm, fill=black] (5.5,-1) rectangle (6.2,1);
\draw[line width=0.01mm, fill=black] (7.2,-1) rectangle (7.9,1);

\draw[line width=0.01mm, fill=black] (8.9,-.5) rectangle (9.6,.5);
\draw[line width=0.01mm, fill=black] (10.6,-.5) rectangle (11.7,.5);
\draw[line width=0.01mm, fill=black] (12.7,-.5) rectangle (13.8,.5);

\draw[line width=0.01mm, fill=black] (14.8,-0.25) rectangle (15.9,0.25);
\draw[line width=0.01mm, fill=black] (16.9,-0.25) rectangle (18.6,0.25);
\draw[line width=0.01mm, fill=black] (19.6,-0.25) rectangle (21.3,0.25);

\draw[line width=0.01mm, fill=black] (22.3,-0.125) rectangle (24,0.125);
\draw[line width=0.01mm, fill=black] (25,-0.125) rectangle (27.5,0.125);
\draw[line width=0.01mm, fill=black] (28.5,-0.125) rectangle (31,0.125);

\draw[line width=0.01mm, fill=black] (32,-.25) rectangle (33.7,.25);
\draw[line width=0.01mm] (33.7,-.25) rectangle (35.4,.25);
\draw[line width=0.01mm, fill=black] (36.4,-.25) rectangle (38.1,.25);
\draw[line width=0.01mm, fill=black] (39.1,-.25) rectangle (40.8,.25);

\draw[line width=0.01mm, fill=black] (41.8,-.5) rectangle (42.9,.5);
\draw[line width=0.01mm] (42.9,-.5) rectangle (44,.5);
\draw[line width=0.01mm, fill=black] (45,-.5) rectangle (46.1,.5);
\draw[line width=0.01mm, fill=black] (47.1,-.5) rectangle (48.2,.5);

\draw[line width=0.01mm, fill=black] (49.2,-1) rectangle (49.9,1);
\draw[line width=0.01mm] (49.9,-1) rectangle (50.6,1);
\draw[line width=0.01mm, fill=black] (51.6,-1) rectangle (52.3,1);
\draw[line width=0.01mm, fill=black] (53.3,-1) rectangle (54,1);

\draw[line width=0.01mm, fill=black] (55,-2) rectangle (55.5,2);
\draw[line width=0.01mm] (55.5,-2) rectangle (56,2);
\draw[line width=0.01mm, fill=black] (57,-2) rectangle (57.5,2);
\draw[line width=0.01mm, fill=black] (58.5,-2) rectangle (59,2);
\draw[line width=0.01mm, fill=black] (60,-2) rectangle (60.1,2);

\draw[line width=1mm, loosely dashed, -latex] (20.45,0.25) to[out=15,in=165] (34.55,.25);
\draw[line width=1mm, loosely dashed, -latex] (13.25,.5) to[out=10,in=170] (43.45,.5);
\draw[line width=1mm, loosely dashed] (7.55,1) to[out=10,in=180] (28.9,2.8);
\draw[line width=1mm, loosely dashed, -latex] (28.9,2.8) to[out=0,in=170] (50.25,1);
\draw[line width=1mm, loosely dashed] (2.75,2) to[out=10,in=180] (29.25,3.9);
\draw[line width=1mm, loosely dashed, -latex] (29.25,3.9) to[out=0,in=170] (55.75,2);

\draw[line width=0.01mm, fill=green] (.2,.4) -- (.2,-.4) -- (.8,0) -- cycle;
\draw[line width=0.01mm, fill=green] (1.7,.4) -- (1.7,-.4) -- (2.3,0) -- cycle;
\draw[line width=0.01mm, fill=red] (3.2,.4) -- (3.2,-.4) -- (3.8,0) -- cycle;

\draw[line width=0.01mm, fill=green] (4.7,.4) -- (4.7,-.4) -- (5.3,0) -- cycle;
\draw[line width=0.01mm, fill=green] (6.4,.4) -- (6.4,-.4) -- (7,0) -- cycle;
\draw[line width=0.01mm, fill=red] (8.1,.4) -- (8.1,-.4) -- (8.7,0) -- cycle;

\draw[line width=0.01mm, fill=green] (9.8,.4) -- (9.8,-.4) -- (10.4,0) -- cycle;
\draw[line width=0.01mm, fill=green] (11.9,.4) -- (11.9,-.4) -- (12.5,0) -- cycle;
\draw[line width=0.01mm, fill=red] (14,.4) -- (14,-.4) -- (14.6,0) -- cycle;

\draw[line width=0.01mm, fill=green] (16.1,.4) -- (16.1,-.4) -- (16.7,0) -- cycle;
\draw[line width=0.01mm, fill=green] (18.8,.4) -- (18.8,-.4) -- (19.4,0) -- cycle;
\draw[line width=0.01mm, fill=red] (21.5,.4) -- (21.5,-.4) -- (22.1,0) -- cycle;

\draw[line width=0.01mm, fill=green] (24.2,.4) -- (24.2,-.4) -- (24.8,0) -- cycle;
\draw[line width=0.01mm, fill=green] (27.7,.4) -- (27.7,-.4) -- (28.3,0) -- cycle;
\draw[line width=0.01mm, fill=blue] (31.2,.4) -- (31.2,-.4) -- (31.8,0) -- cycle;

\draw[line width=0.01mm, fill=green] (35.6,.4) -- (35.6,-.4) -- (36.1,0) -- cycle;
\draw[line width=0.01mm, fill=green] (38.3,.4) -- (38.3,-.4) -- (38.9,0) -- cycle;
\draw[line width=0.01mm, fill=blue] (41,.4) -- (41,-.4) -- (41.6,0) -- cycle;

\draw[line width=0.01mm, fill=green] (44.2,.4) -- (44.2,-.4) -- (44.8,0) -- cycle;
\draw[line width=0.01mm, fill=green] (46.3,.4) -- (46.3,-.4) -- (46.9,0) -- cycle;
\draw[line width=0.01mm, fill=blue] (48.4,.4) -- (48.4,-.4) -- (49,0) -- cycle;

\draw[line width=0.01mm, fill=green] (50.8,.4) -- (50.8,-.4) -- (51.4,0) -- cycle;
\draw[line width=0.01mm, fill=green] (52.5,.4) -- (52.5,-.4) -- (53.1,0) -- cycle;
\draw[line width=0.01mm, fill=blue] (54.2,.4) -- (54.2,-.4) -- (54.8,0) -- cycle;

\draw[line width=0.01mm, fill=green] (56.2,.4) -- (56.2,-.4) -- (56.8,0) -- cycle;
\draw[line width=0.01mm, fill=green] (57.7,.4) -- (57.7,-.4) -- (58.3,0) -- cycle;
\draw[line width=0.01mm] (59.2,.4) -- (59.2,-.4) -- (59.8,0) -- cycle;

\node at (-0.05,-2.6) {\fontsize{30}{30}\selectfont 1};
\node at (1.25,-2.6) {\fontsize{30}{30}\selectfont 32};
\node at (2.75,-2.6) {\fontsize{30}{30}\selectfont 32};

\node at (4.25,-1.6) {\fontsize{30}{30}\selectfont 32};
\node at (5.85,-1.6) {\fontsize{30}{30}\selectfont 64};
\node at (7.55,-1.6) {\fontsize{30}{30}\selectfont 64};

\node at (9.25,-1.1) {\fontsize{30}{30}\selectfont 64};
\node at (11.15,-1.1) {\fontsize{30}{30}\selectfont 128};
\node at (13.25,-1.1) {\fontsize{30}{30}\selectfont 128};

\node at (15.35,-.85) {\fontsize{30}{30}\selectfont 128};
\node at (17.75,-.85) {\fontsize{30}{30}\selectfont 256};
\node at (20.45,-.85) {\fontsize{30}{30}\selectfont 256};

\node at (23.15,-.725) {\fontsize{30}{30}\selectfont 256};
\node at (26.25,-.725) {\fontsize{30}{30}\selectfont 512};
\node at (29.75,-.725) {\fontsize{30}{30}\selectfont 512};

\node at (32.75,-.85) {\fontsize{30}{30}\selectfont 256};
\node at (34.65,-.85) {\fontsize{30}{30}\selectfont 256};
\node at (37.25,-.85) {\fontsize{30}{30}\selectfont 256};
\node at (39.95,-.85) {\fontsize{30}{30}\selectfont 256};

\node at (42.05,-1.1) {\fontsize{30}{30}\selectfont 128};
\node at (43.75,-1.1) {\fontsize{30}{30}\selectfont 128};
\node at (45.55,-1.1) {\fontsize{30}{30}\selectfont 128};
\node at (47.65,-1.1) {\fontsize{30}{30}\selectfont 128};

\node at (49.3,-1.6) {\fontsize{30}{30}\selectfont 64};
\node at (50.5,-1.6) {\fontsize{30}{30}\selectfont 64};
\node at (51.95,-1.6) {\fontsize{30}{30}\selectfont 64};
\node at (53.65,-1.6) {\fontsize{30}{30}\selectfont 64};

\node at (54.95,-2.6) {\fontsize{30}{30}\selectfont 32};
\node at (56.05,-2.6) {\fontsize{30}{30}\selectfont 32};
\node at (57.25,-2.6) {\fontsize{30}{30}\selectfont 32};
\node at (58.75,-2.6) {\fontsize{30}{30}\selectfont 32};
\node at (60.05,-2.6) {\fontsize{30}{30}\selectfont 1};

\draw[line width=0.01mm, fill=black] (3,5) rectangle (3.1,9);
\draw[line width=0.01mm, fill=black] (4.1,5) rectangle (4.6,9);

\draw[line width=0.01mm, fill=black] (5.6,6) rectangle (6.1,8);
\draw[line width=0.01mm, fill=black] (7.1,6) rectangle (7.8,8);

\draw[line width=0.01mm, fill=black] (8.8,6.5) rectangle (9.5,7.5);
\draw[line width=0.01mm, fill=black] (10.5,6.5) rectangle (11.6,7.5);

\draw[line width=0.01mm, fill=black] (12.6,6.75) rectangle (13.7,7.25);
\draw[line width=0.01mm, fill=black] (14.7,6.75) rectangle (16.4,7.25);

\draw[line width=0.01mm, fill=black] (17.4,6.875) rectangle (19.1,7.125);
\draw[line width=0.01mm, fill=black] (20.1,6.875) rectangle (22.6,7.125);

\draw[line width=0.01mm, fill=black] (23.6,6.9375) rectangle (26.1,7.0625);
\draw[line width=0.01mm, fill=black] (27.1,6.9375) rectangle (30.8,7.0625);

\draw[line width=0.01mm, fill=black] (32,6.875) rectangle (34.5,7.125);
\draw[line width=0.01mm, fill=black] (35.5,6.875) rectangle (38,7.125);

\draw[line width=0.01mm, fill=black] (39,6.75) rectangle (40.7,7.25);
\draw[line width=0.01mm, fill=black] (41.7,6.75) rectangle (43.4,7.25);

\draw[line width=0.01mm, fill=black] (44.4,6.5) rectangle (45.5,7.5);
\draw[line width=0.01mm, fill=black] (46.5,6.5) rectangle (47.6,7.5);

\draw[line width=0.01mm, fill=black] (48.6,6) rectangle (49.3,8);
\draw[line width=0.01mm, fill=black] (50.3,6) rectangle (51,8);

\draw[line width=0.01mm, fill=black] (52,5) rectangle (52.5,9);
\draw[line width=0.01mm, fill=black] (53.5,5) rectangle (54,9);
\draw[line width=0.01mm, fill=black] (55,5) rectangle (55.1,9);

\draw[line width=1mm, loosely dotted ] (31,9) -- (31,6);
\node[anchor=west] at (31.2,8.6) {\scalebox{3.4}{Decoder $g$}};
\node[anchor=east] at (30.8,8.6) {\scalebox{3.4}{Encoder $f$}};

\draw[line width=0.01mm, fill=green] (3.3,6.6) -- (3.3,7.4) -- (3.9,7) -- cycle;
\draw[line width=0.01mm, fill=red] (4.8,6.6) -- (4.8,7.4) -- (5.4,7) -- cycle;

\draw[line width=0.01mm, fill=green] (6.3,6.6) -- (6.3,7.4) -- (6.9,7) -- cycle;
\draw[line width=0.01mm, fill=red] (8,6.6) -- (8,7.4) -- (8.6,7) -- cycle;

\draw[line width=0.01mm, fill=green] (9.7,6.6) -- (9.7,7.4) -- (10.3,7) -- cycle;
\draw[line width=0.01mm, fill=red] (11.8,6.6) -- (11.8,7.4) -- (12.4,7) -- cycle;

\draw[line width=0.01mm, fill=green] (13.9,6.6) -- (13.9,7.4) -- (14.5,7) -- cycle;
\draw[line width=0.01mm, fill=red] (16.6,6.6) -- (16.6,7.4) -- (17.2,7) -- cycle;

\draw[line width=0.01mm, fill=green] (19.3,6.6) -- (19.3,7.4) -- (19.9,7) -- cycle;
\draw[line width=0.01mm, fill=red] (22.8,6.6) -- (22.8,7.4) -- (23.4,7) -- cycle;

\draw[line width=0.01mm, fill=green] (26.3,6.6) -- (26.3,7.4) -- (26.9,7) -- cycle;
\draw[line width=0.01mm, fill=blue] (31.2,6.6) -- (31.2,7.4) -- (31.8,7) -- cycle;

\draw[line width=0.01mm, fill=green] (34.7,6.6) -- (34.7,7.4) -- (35.3,7) -- cycle;
\draw[line width=0.01mm, fill=blue] (38.2,6.6) -- (38.2,7.4) -- (38.8,7) -- cycle;

\draw[line width=0.01mm, fill=green] (40.9,6.6) -- (40.9,7.4) -- (41.5,7) -- cycle;
\draw[line width=0.01mm, fill=blue] (43.6,6.6) -- (43.6,7.4) -- (44.2,7) -- cycle;

\draw[line width=0.01mm, fill=green] (45.7,6.6) -- (45.7,7.4) -- (46.3,7) -- cycle;
\draw[line width=0.01mm, fill=blue] (47.8,6.6) -- (47.8,7.4) -- (48.4,7) -- cycle;

\draw[line width=0.01mm, fill=green] (49.5,6.6) -- (49.5,7.4) -- (50.1,7) -- cycle;
\draw[line width=0.01mm, fill=blue] (51.2,6.6) -- (51.2,7.4) -- (51.8,7) -- cycle;

\draw[line width=0.01mm, fill=green] (52.7,6.6) -- (52.7,7.4) -- (53.3,7) -- cycle;
\draw[line width=0.01mm] (54.2,6.6) -- (54.2,7.4) -- (54.8,7) -- cycle;

\node at (3.05,4.4) {\fontsize{30}{30}\selectfont 1};
\node at (4.35,4.4) {\fontsize{30}{30}\selectfont 16};

\node at (5.85,5.4) {\fontsize{30}{30}\selectfont 16};
\node at (7.45,5.4) {\fontsize{30}{30}\selectfont 32};

\node at (9.15,5.9) {\fontsize{30}{30}\selectfont 32};
\node at (11.05,5.9) {\fontsize{30}{30}\selectfont 64};

\node at (13.15,6.15) {\fontsize{30}{30}\selectfont 64};
\node at (15.55,6.15) {\fontsize{30}{30}\selectfont 128};

\node at (18.25,6.275) {\fontsize{30}{30}\selectfont 128};
\node at (21.35,6.275) {\fontsize{30}{30}\selectfont 256};

\node at (24.85,6.3375) {\fontsize{30}{30}\selectfont 256};
\node at (28.95,6.3375) {\fontsize{30}{30}\selectfont 512};

\node at (33.25,6.275) {\fontsize{30}{30}\selectfont 256};
\node at (36.75,6.275) {\fontsize{30}{30}\selectfont 256};

\node at (39.85,6.15) {\fontsize{30}{30}\selectfont 128};
\node at (42.45,6.15) {\fontsize{30}{30}\selectfont 128};

\node at (44.95,5.9) {\fontsize{30}{30}\selectfont 64};
\node at (47.05,5.9) {\fontsize{30}{30}\selectfont 64};

\node at (48.95,5.4) {\fontsize{30}{30}\selectfont 32};
\node at (50.65,5.4) {\fontsize{30}{30}\selectfont 32};

\node at (52.25,4.4) {\fontsize{30}{30}\selectfont 16};
\node at (53.75,4.4) {\fontsize{30}{30}\selectfont 16};
\node at (55.05,4.4) {\fontsize{30}{30}\selectfont 1};

\draw[line width=0.01mm, fill=black] (3.1,-5) rectangle (3.2,-9);
\draw[line width=0.01mm, fill=black] (4.2,-6) rectangle (4.7,-8);
\draw[line width=0.01mm, fill=black] (5.7,-6.5) rectangle (6.8,-7.5);
\draw[line width=0.01mm, fill=black] (7.8,-6.75) rectangle (9.5,-7.25);
\draw[line width=0.01mm, fill=black] (10.5,-6.75) rectangle (13,-7.25);
\draw[line width=0.01mm, fill=black] (14,-6.75) rectangle (14.1,-7.25);

\draw[line width=0.01mm, fill=orange] (3.4,-6.6) -- (3.4,-7.4) -- (4,-7) -- cycle;
\draw[line width=0.01mm, fill=orange] (4.9,-6.6) -- (4.9,-7.4) -- (5.5,-7) -- cycle;
\draw[line width=0.01mm, fill=orange] (7,-6.6) -- (7,-7.4) -- (7.6,-7) -- cycle;
\draw[line width=0.01mm, fill=magenta] (9.7,-6.6) -- (9.7,-7.4) -- (10.3,-7) -- cycle;
\draw[line width=0.01mm, fill=violet] (13.2,-6.6) -- (13.2,-7.4) -- (13.8,-7) -- cycle;

\node at (3.15,-9.6) {\fontsize{30}{30}\selectfont 1};
\node at (4.45,-8.6) {\fontsize{30}{30}\selectfont 32};
\node at (6.25,-8.1) {\fontsize{30}{30}\selectfont 64};
\node at (8.65,-7.85) {\fontsize{30}{30}\selectfont 128};
\node at (11.75,-7.85) {\fontsize{30}{30}\selectfont 256};
\node at (14.05,-7.85) {\fontsize{30}{30}\selectfont 1};

\node at (30.1,5) {\scalebox{3.4}{Convolutional Auto-Encoder (CAE)}};

\node at (30.1,-2.5) {\scalebox{3.4}{Generator network $G$ based on UNet}};

\node at (8.6,-10.55) {\scalebox{3.4}{Discriminator network $D$}};

\draw[loosely dashdotted] (55,-4) rectangle (17,-11);

\draw[line width=0.01mm, fill=green] (17.5,-4.5) -- (17.5,-5.3) -- (18.1,-4.9) -- cycle;
\draw[line width=0.01mm, fill=red] (17.5,-5.8) -- (17.5,-6.6) -- (18.1,-6.2) -- cycle;
\draw[line width=0.01mm, fill=blue] (17.5,-7.1) -- (17.5,-7.9) -- (18.1,-7.5) -- cycle;
\draw[line width=0.01mm] (17.5,-8.4) -- (17.5,-9.2) -- (18.1,-8.8) -- cycle;
\draw[line width=1mm,loosely dashed, -latex] (17.5,-10.1) -- (18.1,-10.1);

\node[anchor=west] at (18.6,-4.9) {\fontsize{30}{30}\selectfont conv 3$\times$3, batch normalization, ReLU};
\node[anchor=west] at (18.6,-6.2) {\fontsize{30}{30}\selectfont max-pooling 2$\times$2};
\node[anchor=west] at (18.6,-7.5) {\fontsize{30}{30}\selectfont up-conv 2$\times$2};
\node[anchor=west] at (18.6,-8.8) {\fontsize{30}{30}\selectfont conv 3$\times$3, sigmoid};
\node[anchor=west] at (18.6,-10.1) {\fontsize{30}{30}\selectfont copy and concatenate};

\draw[line width=0.01mm, fill=orange] (35.9,-4.5) -- (35.9,-5.3) -- (36.5,-4.9) -- cycle; 
\draw[line width=0.01mm, fill=magenta] (35.9,-7.1) -- (35.9,-7.9) -- (36.5,-7.5) -- cycle; 
\draw[line width=0.01mm, fill=violet] (35.9,-9.7) -- (35.9,-10.5) -- (36.5,-10.1) -- cycle; 

\node[anchor=west] at (37,-4.9) {\fontsize{30}{30}\selectfont conv 4$\times$4, strides 2, batch normalization,};
\node[anchor=west] at (37,-6.2) {\fontsize{30}{30}\selectfont ReLU};
\node[anchor=west] at (37,-7.5) {\fontsize{30}{30}\selectfont conv 4$\times$4, strides 1, batch normalization,};
\node[anchor=west] at (37,-8.8) {\fontsize{30}{30}\selectfont ReLU};
\node[anchor=west] at (37,-10.1) {\fontsize{30}{30}\selectfont conv 4$\times$4, strides 1, sigmoid};

\end{tikzpicture}
\end{adjustbox}
\caption{Proposed architectures: CAE comprising encoder $f$, decoder $g$ (top), generator $G$ (middle), and discriminator $D$ (down).}
\label{fig:fig_archi}
\end{figure*}

Pediatric shoulder musculo-skeletal disorders such as impingement syndrome, labral tears and shoulder dislocations have debilitating impact on child's growth and affect their daily living activities \cite{bishop_pediatric_2005}. Many of these pathologies include deformation of scapular bone along with abnormality of surrounding musculature. Hence, it is essential to understand the changed morphology in pathological population and thus accurate and fast scapular segmentation becomes a necessity. 


Works dedicated to scapula segmentation in MR images are currently oriented towards Convolutional Neural Networks (CNN) \cite{he_effective_2019}. Although, the work of He et al. \cite{he_effective_2019} uses VNet to include 3D spatial information, it does not incorporate additional information such as shape priors within the CNN, which are needed for robust segmentation. Moreover, 3D networks are computationally expensive and require high GPU memory consumption compared to their 2D counterparts.

To improve the accuracy of segmentation, recent research aims at incorporating shape priors into the network. Oktay et al. \cite{oktay_anatomically_2018} penalized the deviation of the predicted segmentation mask from a learnt shape model, by using an extended UNet with an encoder pre-trained for shape regularization. The segmentation network was trained using a novel loss function combining both cross-entropy and shape regularization losses. Furthermore, Pham et al. \cite{pham_deep_2019} proposed an imitating encoder to incorporate anatomical priors. Another recent trend consists in employing conditional adversarial frameworks for medical image segmentation. Following image-to-image translation approaches \cite{isola_image--image_2017}, Singh et al. \cite{singh_conditional_2018} proposed a model based on a generator and a discriminator networks trained sequentially. By including the adversarial score, the generator was guided towards realistic segmentations.

In this context, the purpose of this work was to incorporate anatomical priors into a conditional adversarial framework, given a limited amount of heterogeneous annotated images. Specifically, we combined shape priors and conditional neural networks, to encourage the model to follow the global anatomical properties in terms of shape and location and to enforce the delineation to be as realistic as possible. We demonstrated the effectiveness of such a combination for scapula MR image segmentation, using a limited number of pediatric shoulder examinations.


\vspace{-.5em}

\section{Method}
\label{sec:methodology}

\begin{table*}[ht!]
\centering
    \begin{tabular}{||c || c c c c c||} 
     \hline
     Methods & Dice & Sensitivity & Specificity & Jaccard & Hausdorff Dist. \\ 
     \hline\hline
     UNet \cite{ronneberger_u-net:_2015} & 79.21$\pm$15.82 & 75.89$\pm$21.59 & \textbf{99.93$\pm$0.01} & 68.01$\pm$18.69 & 24.35$\pm$20.84\\ 
     CAE-UNet \cite{oktay_anatomically_2018} & 80.52$\pm$14.01 & 77.48$\pm$18.32 & \textbf{99.93$\pm$0.01} & 69.27$\pm$16.23 & 29.63$\pm$28.18 \\
     cGAN-UNet \cite{singh_conditional_2018} & 80.69$\pm$13.12 & 78.78$\pm$18.19 & 99.92$\pm$0.01 & 69.30$\pm$15.27 &   24.07$\pm$22.27 \\
     Our method & \textbf{82.19$\pm$9.96} & \textbf{80.26$\pm$14.01} & 99.91$\pm$0.01 &  \textbf{70.87$\pm$13.17} &  \textbf{19.31$\pm$16.49} \\
     \hline
    \end{tabular}
\caption{Leave-one-out quantitative assessment of UNet \cite{ronneberger_u-net:_2015}, CAE-UNet \cite{oktay_anatomically_2018}, cGAN-UNet \cite{singh_conditional_2018} and our method in Dice, sensitivity, specificity, Jaccard scores ($\%$) and Haussdorff distance. (mm). Best results are in bold.}
\label{table:1}
\end{table*}

Our approach combined a non-linear representation of the shape with an adversarial contribution to enhance the segmentation framework (Fig.\ref{fig:fig_framework}). The proposed framework consisted of three modules: a generator $G$ based on UNet, a discriminator $D$ and a Convolutional Auto-Encoder (CAE). UNet learned the intrinsic features to generate a binary mask. The discriminator assessed if a given binary mask was likely to be a realistic segmentation or not. Lastly, a CAE was trained beforehand on the binary mask to learn a non-linear representation of the scapula shape.
\vspace{-.75em}

\subsection{Baseline UNet}

Let $x$ be the observed intensity image, and $y$ the corresponding image of class labels. The mapping between intensities and labels 
$\begin{array}{ccccccccc}
G & : & X & \to & \mathcal{L} & ,& x & \mapsto & G(x) \\
\end{array}$ 
was learnt by optimizing the loss function through stochastic gradient descent. In baseline UNet, the loss function was determined using: $L_{Dice}(y,G(x)) = 1 - \dfrac{2|y\cap G(x)|}{|y|+|G(x)|}$ where $|.|$ was the total sum of the pixel values of a given binary image. The generator consisted of several convolutional encoding and deconvolutional decoding layers as well as skip connections between them to retain the localization information from input features (Fig.\ref{fig:fig_archi}).

\vspace{-.75em}

\subsection{Incorporating shape priors}
\label{ssec:CAE}

An auto-encoder is a neural network which can learn the latent representation of the data from which the original data can be reconstructed. The encoder $f$ maps the input to a low dimensional manifold, whereas the decoder $g$ reconstructs the original input from the compact representation. During training, the auto-encoder learns the most salient features of the training data. Similar to Oktay et al.'s \cite{oktay_anatomically_2018} approach, we used a CAE (Fig.\ref{fig:fig_archi}) to find an embedding in latent space, that encoded the anatomical priors of the scapula arising from groundtruth annotations. The learning scheme of the CAE minimized a Dice loss function, $L_{CAE}(f,g) = \mathop{{}\mathbb{E}}_{y} [L_{Dice}(y, g(f(y)))]$, penalizing the reconstruction from being dissimilar to original segmentation.


During the segmentation model training, both the prediction and the groundtruth labels were projected on lower dimensional manifold using the encoder component of the CAE with fixed weights (Fig.\ref{fig:fig_archi}). We defined the shape regularization loss as : $L_h = \mathop{{}\mathbb{E}}_{x, y} [ (f( G(x)) - f(y))^{2} ]$.

\vspace{-.75em}

\subsection{Combining shape priors and adversarial networks}
\label{ssec:GAN}

We adapted the conditional generative adversarial network image-to-image translation pipeline \cite{singh_conditional_2018} to our problem. The generator $G$ generated binary mask while the discriminator $D$ (Fig.\ref{fig:fig_archi}) assessed if a given binary mask is fake or not. We defined the adversarial loss as : $L_a = \mathop{{}\mathbb{E}}_{x} [-\log(D(x,G(x)))]$. The adversarial loss could be explained as a similarity measure between generated outputs and groundtruth masks. This score would improve the capability of $G$ to provide correct segmentation. 




\begin {figure*}[t]
\centering
\begin{adjustbox}{width=\textwidth}
\begin{tikzpicture}
\begin{scope}[spy using outlines=
      {circle, magnification=2, size=.4cm, connect spies, rounded corners}]

\node[inner sep=0pt] (unet-1) at (0,0)
    {\includegraphics[width=.08\textwidth]{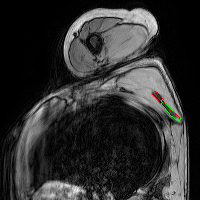}};
\node[inner sep=0pt] (cgan-1) at (1.5,0)
    {\includegraphics[width=.08\textwidth]{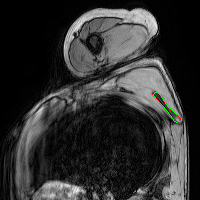}};
\node[inner sep=0pt] (cae-1) at (3,0)
    {\includegraphics[width=.08\textwidth]{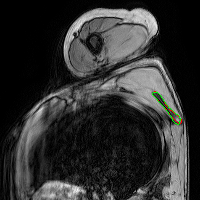}};
\node[inner sep=0pt] (our-1) at (4.5,0)
    {\includegraphics[width=.08\textwidth]{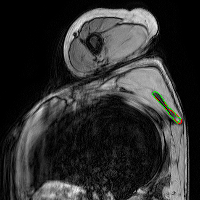}};

\node[inner sep=0pt] (unet-2) at (0,-1.5)
    {\includegraphics[width=.08\textwidth]{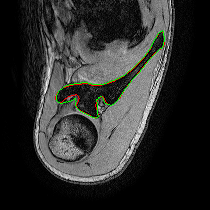}};
\node[inner sep=0pt] (cgan-2) at (1.5,-1.5)
    {\includegraphics[width=.08\textwidth]{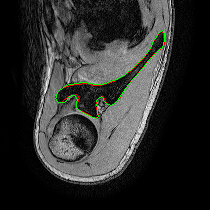}};
\node[inner sep=0pt] (cae-2) at (3,-1.5)
    {\includegraphics[width=.08\textwidth]{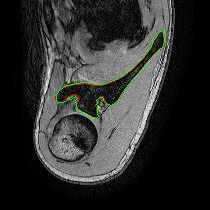}};
\node[inner sep=0pt] (our-2) at (4.5,-1.5)
    {\includegraphics[width=.08\textwidth]{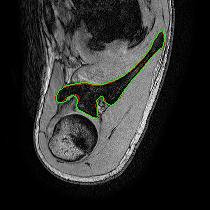}};

\spy [orange] on (.42,0) in node [left] at (-.27,.47);
\spy [orange] on (1.92,0) in node [left] at (1.23,.47);
\spy [orange] on (3.42,0) in node [left] at (2.73,.47);
\spy [orange] on (4.92,0) in node [left] at (4.23,.47);

\spy [orange] on (-.2,-1.46) in node [left] at (.67,-1.97);
\spy [orange] on (1.3,-1.46) in node [left] at (2.17,-1.97);
\spy [orange] on (2.8,-1.46) in node [left] at (3.67,-1.97);
\spy [orange] on (4.3,-1.46) in node [left] at (5.17,-1.97);

\end{scope}


\node at (0, .8) {\fontsize{3}{3}\selectfont UNet};
\node at (1.5, .8) {\fontsize{3}{3}\selectfont CAE-UNet};
\node at (3, .8) {\fontsize{3}{3}\selectfont cGAN-UNet};
\node at (4.5, .8) {\fontsize{3}{3}\selectfont Proposed method};

\end{tikzpicture}
\end{adjustbox}
\caption{Automatic pathological segmentation of scapula using UNet \cite{ronneberger_u-net:_2015}, CAE-UNet \cite{oktay_anatomically_2018}, cGAN-UNet \cite{singh_conditional_2018} and proposed method. Groundtruth and estimated delineations are in red and green respectively. In the first comparison, UNet under-segments the bone area, whereas other approaches achieve more accurate scapula delineations. In the second example, the proposed method captures more complex bone shape and subtle contours compared to other architectures.}
\label{fig:fig_comparison}
\end{figure*}
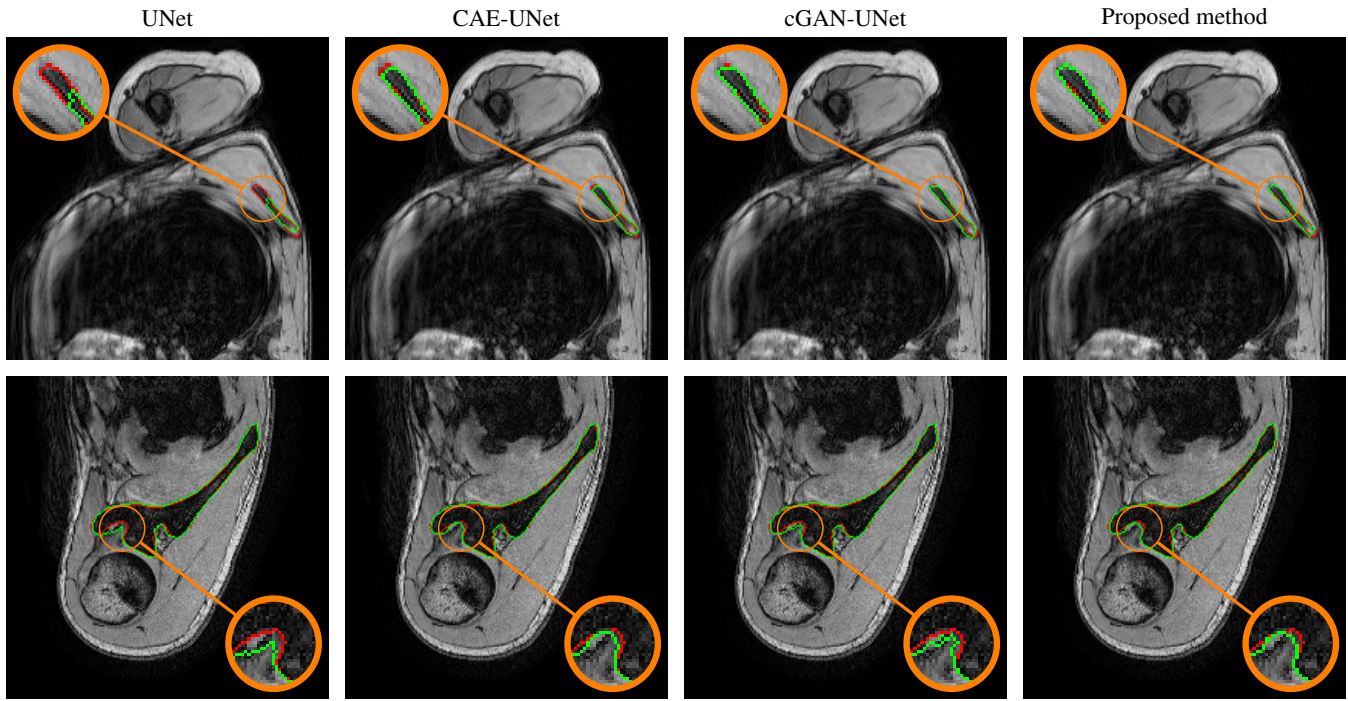


Finally, we defined the proposed loss function of $G$ as: $L_{G}(G,D,f) =  \mathop{{}\mathbb{E}}_{x, y} [L_{Dice}(y,G(x))]  + \lambda_{1} L_{a}  + \lambda_2 L_{h}$, with empirical weighting factors $\lambda_1$ and $\lambda_2$. The proper optimization of $G$ was enforced by each term of the loss function. The Dice loss promoted a coarse prediction of the segmentation while the adversarial term $L_{a}$ encouraged a precise prediction of the mask outline. Simultaneously, the latent loss $L_{h}$ promoted globally consistent and plausible shape delineations. Additionally, the loss function of $D$ was defined as : $L_{D}(G,D) = \mathop{{}\mathbb{E}}_{x, y} [-\log(D(x), y)] +   \mathop{{}\mathbb{E}}_{x, y} [-\log(1-  D(x,G(x)))]$.





As post-processing step, the obtained 2D binary slices were stacked to form a 3D volume and the largest connected set was selected as the final 3D predicted scapular mask.

\vspace{-.5em}

\section{Experiments}
\label{sec:experiments}

\subsection{Imaging scapula dataset}
\label{ssec:dataset}

MR images of fifteen pediatric shoulder complex were acquired using a 3T Philips Achieva (Philips Healthcare, Best, The Netherlands) scanner. An enhanced T1-weighted  high resolution isotropic volume examination (eTHRIVE) sequence was used. All the images were annotated by an expert (radiologist, 12 years of experience) to get groundtruth segmentation. Image size and quality varied for each subject, all axial slices were downsampled to 256$\times$256 pixels to reduce computation time.


\subsection{Segmentation assessment}
\label{ssec:assessment}

To assess the performance of the proposed architecture, the accuracy of the 3D scapula segmentation was quantified based on Dice $(\tfrac{2 TP}{2TP+FP+FN})$, sensitivity $(\tfrac{TP}{TP+FN})$, specificity $(\tfrac{TN}{TN+FP})$ and Jaccard Index $(\tfrac{TP}{TP+FP+FN})$ scores, where TP, FP, TN and FN were the number of true or false positive and negative voxels. The quality of the segmentation was also evaluated using Hausdorff distance $HD(A,B)=\max(h(A,B),h(B,A))$, where $h(A,B) = \max_{a \in A} \min_{b \in B} ||a-b||$ and $A$ and $B$  were respectively the set of non-zero voxels in labels images. These scores provided an assessment of the models' ability to generate the same contours as those produced manually. 

\vspace{-.75em}
\subsection{Training details}

As a first step, we optimized the CAE using Dice loss. We explored various learning rates for Adam optimizer. Adam with initial learning rate $0.01$, batch size 32 and 20 epochs was found to be the best combination. Then, as a second step, the optimization of $G$ and $D$ was done sequentially with the weights of the encoder $f$ fixed. We experimented different hyperparameters. Adam with initial learning rate $0.0001$, batch size 32 and 20 epochs suited the best. The adversarial loss weighting factor $\lambda_1 = 0.01$ and the latent loss weighting factor $\lambda_2 = 0.0001$ were found to be optimal. All networks were trained using data augmentation since the amount of available training data was limited. Training 2D axial slices underwent random scaling, rotation, shifting on both directions, to teach the networks the desired invariance, covariance and robustness characteristics. Models were implemented using Keras and trained using a Nvidia RTX 2080 Ti GPU with 11 GB of Video RAM. 


\vspace{-.75em}

\subsection{Results}
\label{ssec:results}

The proposed scapula segmentation method was compared with other deep end-to-end architectures and evaluated quantitatively. The other three methods were UNet \cite{ronneberger_u-net:_2015}, cGAN-UNet \cite{singh_conditional_2018} and CAE-UNet \cite{oktay_anatomically_2018}. The same UNet architecture was used for all 4 methods. All the architectures were trained from scratch. Experiments were performed in a leave-one-out fashion to assess the generality of each model. 


From the quantitative assessment comparison (Tab.\ref{table:1}), our method outperformed the compared state-of-the-art methods with regards to Dice, Sensitivity, Jaccard index and Hausdorff distance. Indeed, our method had highest average Dice with 82.19 compared to 79.21, 80.52 and 80.69 for UNet, CAE-UNet and cGAN-UNet respectively. Specificity was excellent for all architectures ranging from 99.91 to 99.93. We assumed that the variability in our results was due to the heterogeneity of our database. For one subject, the difference between bone intensity and muscle intensity was smaller than in the rest of our dataset which resulted in poor segmentation. The segmentation results revealed the importance of combining both shape priors and adversarial terms during training.





\vspace{-.5em}

\section{Conclusion}
\label{sec:conclusion}

In this work, we presented a  deep learning architecture with promising results for scapula MR image segmentation. Our framework employed both an auto-encoder and conditional adversarial network as regularizers to train a convolutional encoder-decoder model. The proposed framework made predictions that were in agreement with the shape prior while the adversarial score encouraged realistic shape. Future work is aimed at improving our framework to detect other anatomical structures such as the humerus or the deltoid. These structures will then be integrated into a more comprehensive diagnostic tool to compute the severity of shoulder pathologies.



%
%
%



\vspace{-.5em}

\bibliographystyle{ieeetr}
\small{
\bibliography{ISBI2020}}
\end{document}